\documentclass[%
 preprint,
 superscriptaddress,
 reprint, twocolumn,
 amsmath,amssymb,
 aps,
 prl,
]{revtex4-2}

\usepackage{graphicx}
\usepackage{dcolumn}
\usepackage{bm}
\usepackage{physics}
\usepackage[colorlinks,linkcolor=blue,anchorcolor=blue,citecolor=blue]{hyperref}

\usepackage{color}

\begin{document}


\title{Decisive role of electron-phonon coupling for phonon and electron instabilities in transition metal dichalcogenides}

\author{Zishen~Wang}
\thanks{These authors contributed equally.}

\affiliation{Department of Physics, National University of Singapore, 117542 Singapore, Singapore}
\affiliation{Centre for Advanced 2D Materials, National University of Singapore, 117546 Singapore, Singapore}

\author{Chuan~Chen}
\thanks{These authors contributed equally.}

\affiliation{
Institute for Advanced Study, Tsinghua University,
100084 Beijing, China
}
\affiliation{
Max-Planck Institute for the Physics of Complex Systems,
01187 Dresden, Germany}

\author{Jinchao Mo}
\affiliation{Department of Physics, National University of Singapore, 117542 Singapore, Singapore}

\author{Jun Zhou}
\email{zhou\_jun@imre.a-star.edu.sg}
\affiliation{Institute of Materials Research \& Engineering, A*STAR (Agency for Science, Technology and Research), 138634 Singapore, Singapore.}

\author{Kian Ping Loh}
\email{chmlohkp@nus.edu.sg}
\affiliation{Centre for Advanced 2D Materials, National University of Singapore, 117546 Singapore, Singapore}
\affiliation{Department of Chemistry, National University of Singapore, 117543 Singapore, Singapore}

\author{Yuan Ping Feng}
\email{phyfyp@nus.edu.sg}
\affiliation{Department of Physics, National University of Singapore, 117542 Singapore, Singapore}
\affiliation{Centre for Advanced 2D Materials, National University of Singapore, 117546 Singapore, Singapore}

\begin{abstract}
The origin of the charge density wave (CDW) in transition metal dichalcognides has been in hot debate and no conclusive agreement has been reached. Here, we propose an \textit{ab-initio} framework for an accurate description of both Fermi surface nesting and electron-phonon coupling (EPC) and systematically investigate their roles in the formation of CDW. Using monolayer 1H-NbSe$_2$ and 1T-VTe$_2$ as representative examples, we show that it is the momentum-dependent EPC softens the phonon frequencies, which become imaginary (phonon instabilities) at CDW vectors (indicating CDW formation). Besides, the distribution of the CDW gap opening (electron instabilities) can be correctly predicted only if EPC is included in the mean-field model. These results emphasize the decisive role of EPC in the CDW formation. Our analytical process is general and can be applied to other CDW systems.

\end{abstract}

\maketitle

The formation of charge density wave (CDW) is a spontaneous symmetry breaking process with periodic charge density modulations and lattice distortions below a critical temperature ($T_{CDW}$) \cite{gruner1988dynamics,rossnagel2011origin}. However, the origin of CDW is a long-standing problem, which attracts broad research interest in condensed matter physics \cite{zhu2017misconceptions,johannes2008fermi,weber2011extended,zhu2015classification}. The first mechanism, Fermi surface nesting (FSN), was originated from Peierls’ model of an ideal one-dimensional (1D) metal atomic chain. FSN relates to an elastic electronic scattering at the Fermi surface \cite{peierls1955quantum}. The zero-energy electronic excitations screen the phonon vibration at the CDW vector \textbf{Q}, inducing an abrupt phonon softening (known as Kohn anomaly) \cite{kohn1959image}. However, the extension of FSN to real materials is not ideal. With a few exceptions, FSN is argued to have limited power in inducing CDW distortions in higher-dimensional systems \cite{johannes2008fermi,weber2011extended,zhu2015classification,diego2021van}.

Instead, momentum-dependent electron-phonon coupling (\textbf{q}-EPC), which involves an inelastic electronic scattering mediated by a phonon, is argued to be more prevailing in higher-dimensional systems \cite{johannes2008fermi,weber2011extended,zhu2015classification,diego2021van}. In an electron-phonon interaction dominant system, the electron field can be integrated out as a perturbation to the free phonon field, softening phonon frequencies from their bare values [see SI-III for details]. Thus, both FSN and \textbf{q}-EPC may soften phonons to imaginary values (phonon instabilities) and induce CDW distortions. However, quantitative studies of \textbf{q}-EPC are rare, and the reported method to directly obtain \textbf{q}-EPC is a tight-binding model by merely using the electronic band structure \cite{varma1979electron,flicker2015charge,flicker2016charge}. Besides, such a semi-empirical method is hard to give rigorous results compared to the first-principles calculations. Therefore, it is imperative to explore a general method to accurately describe \textbf{q}-EPC in the CDW materials. 

Another important feature of a CDW is the band gap opening (electron instabilities) accompanied by the CDW distortions, which lowers the energy of the system \cite{rossnagel2011origin,johannes2008fermi,wang2021controllable}. The location of the CDW gap in the Brillouin zone (BZ) can be identified accurately by a band unfolding scheme \cite{zheng2018first,lian2018unveiling,chen2018unique}, which however requires the prior knowledge of the corresponding CDW structure and cannot provide insight into the underlying mechanism. Thus, the driving force of the CDW gap is still elusive. A convenient method that can give physical insight into the gap opening is needed.

In this letter, we use monolayer 1H-NbSe$_2$ and 1T-VTe$_2$ (in short NbSe$_2$ and VTe$_2$) as representative examples for the most common high-symmetry phases of the transition metal dichalcogenides (TMDs). Another reason for such a choice is the mechanisms of their CDWs are considered to be different. The electron-phonon coupling (EPC) has been shown to be dominant in NbSe$_2$ \cite{lian2018unveiling,zheng2019electron,calandra2009effect}, while FSN in VTe$_2$ seems to be substantial, which leads to a peak in the static Lindhard susceptibility \cite{wang2019evidence,sugawara2019monolayer}. Thus, we conduct a comprehensive study of their CDW properties from both phononic and electronic perspectives. In the phononic part, we obtain the accurate \textbf{q}-EPC from the fully first-principles calculations, which allows us to make a straightforward comparison between the contributions from FSN and \textbf{q}-EPC in the CDW formation. Interestingly, the \textbf{q}-EPC is shown to play the dominant role in designating the CDW vectors for both NbSe$_2$ and VTe$_2$. While in the electronic part, we find the rigorous EPC matrix elements are the key to predict the distribution of the CDW gaps by our mean-field model. In addition, we argue that our analysis process, besides its conciseness and accurateness, should be a general framework, which can be easily exploited in other CDW systems.

\begin{figure} [!htb] 
\centering  
\includegraphics[width=8cm]{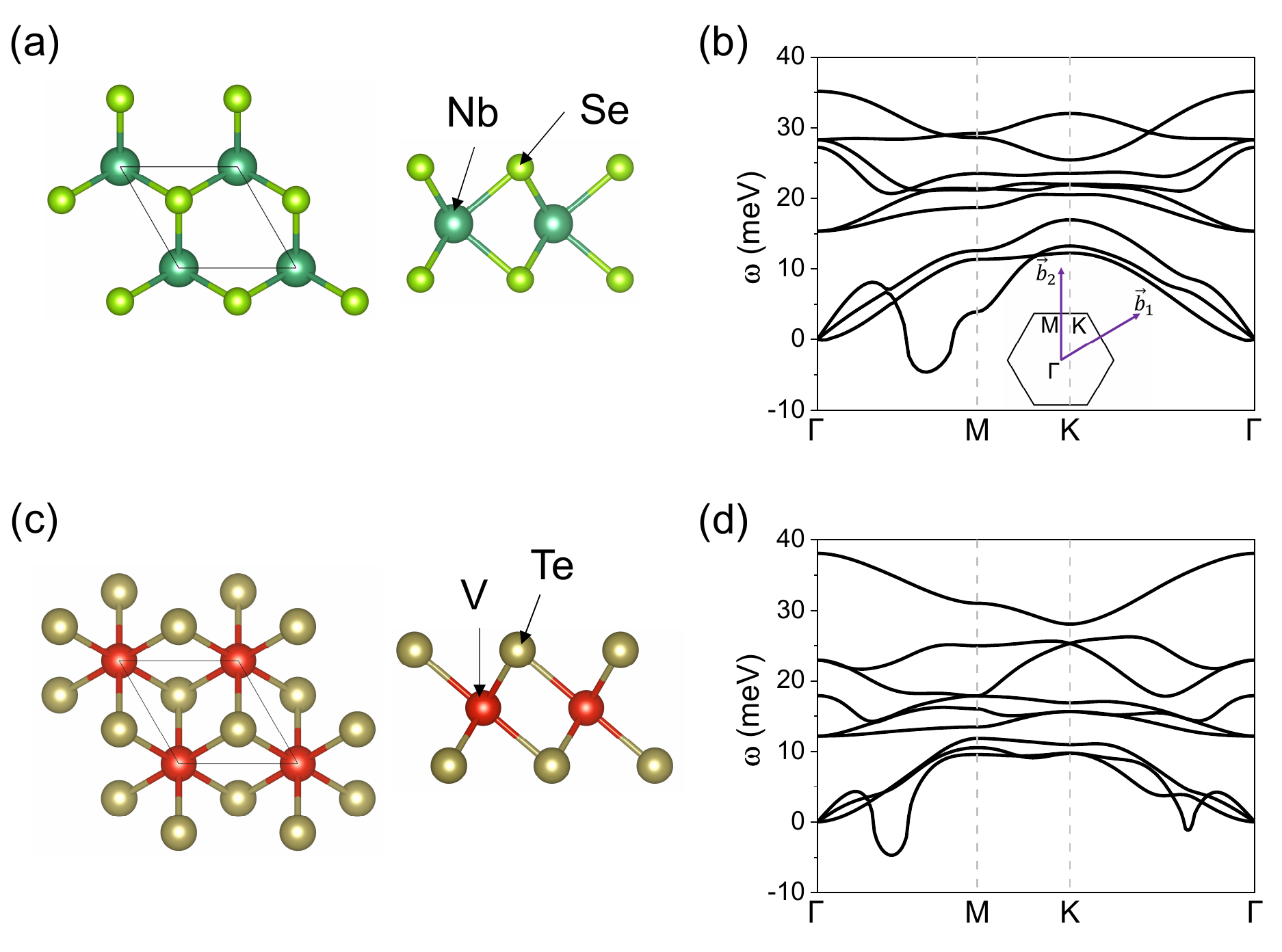}  
\caption{(a) Top and side views of the crystal structure, (b) phonon dispersion of monolayer 1H-NbSe$_2$. (c, d) Same as panel (a, b) but for monolayer 1T-VTe$_2$. The insert figure in (b) displays the first BZ with reciprocal lattice vectors. 
}  
\end{figure}

\textit{Phonon instabilities and CDW mechanism.}--TMDs have layered structures, in which transition metal atoms are sandwiched by chalcogen atoms, forming trigonal prisms (a 1H phase) or octahedrons (a 1T phase). The structures of 1H-NbSe$_2$ and 1T-VTe$_2$ are shown in Figs. 1(a) and 1(c), respectively. As the phonon spectra for NbSe$_2$ and VTe$_2$ shown in Figs. 1(b) and 1(d), their longitudinal acoustic (LA) phonon modes collapse at the CDW vectors, which triple the unit cell for NbSe$_2$ with $\mathbf{Q^H}=2/3\Gamma M$, and quadruple the unit cell for VTe$_2$ with $\mathbf{Q^T}=1/2\Gamma M$, in line with the CDW supercells obtained from experiments \cite{wang2019evidence,ugeda2016characterization,nakata2018anisotropic,coelho2019monolayer}.

To account for the driving force of the phonon softening, the relationship between the softened phonon frequency $\omega_\mathbf{q}$ and its bare phonon frequency $\Omega_\mathbf{q}$ is described under the random phase approximation (RPA) [See more details in SI-III]: 
\begin{equation}
    \omega_\mathbf{q}^2=\Omega_\mathbf{q}^2-2\Omega_\mathbf{q}\chi_\mathbf{q}
\end{equation}

\noindent where $\chi_\mathbf{q}$ is the generalized static electronic susceptibility, including both contributions from FSN and EPC, which is given by:
\begin{equation}
    \chi_\mathbf{q}=\sum_{\mathbf{k}}{\left|g_{\mathbf{k},\mathbf{k}+\mathbf{q}}\right|^2\frac{f(\varepsilon_\mathbf{k})-f(\varepsilon_{\mathbf{k}+\mathbf{q}})}{\varepsilon_{\mathbf{k}+\mathbf{q}}-\varepsilon_\mathbf{k}}}
\end{equation}

\noindent where $f(\varepsilon)$ is the Fermi-Dirac function of the eigenvalue $\varepsilon$, $g_{\mathbf{k},\mathbf{k}+\mathbf{q}}$ is the electron-phonon coupling matrix element that couples electronic states \textbf{k} and \textbf{k+q} with a phonon of momentum \textbf{q}. According to eq. (1), the ordering vector is estimated from the maximum of $\chi_\mathbf{q}$ \cite{flicker2015charge,flicker2016charge}. The $\chi_\mathbf{q}$ often reduces to the static Lindhard susceptibility ${\chi'}_\mathbf{q}$ under the constant matrix element approximation ($|g_{\mathbf{k},\mathbf{k}+\mathbf{q}}|=1$):

\begin{equation}
    {\chi'}_\mathbf{q}=\sum_{\mathbf{k}}\frac{f(\varepsilon_\mathbf{k})-f(\varepsilon_{\mathbf{k}+\mathbf{q}})}{\varepsilon_{\mathbf{k}+\mathbf{q}}-\varepsilon_\mathbf{k}}
\end{equation}

\noindent which is a pure electron effect, and its peak reflects the electronic instability by FSN \cite{johannes2008fermi}. Similarly, $\chi_\mathbf{q}$ can reduce to the \textbf{q}-EPC ${\bar{g}}_\mathbf{q}$ under the “constant fraction” approximation ($\frac{f(\varepsilon_\mathbf{k})-f(\varepsilon_{\mathbf{k}+\mathbf{q}})}{\varepsilon_{\mathbf{k}+\mathbf{q}}-\varepsilon_\mathbf{k}}=1$):

\begin{equation}
    {\bar{g}}_\mathbf{q}=\sum_{\mathbf{k}}\left|g_{\mathbf{k},\mathbf{k}+\mathbf{q}}\right|^2
\end{equation}

\noindent which reflects a pure EPC effect. Focusing on the low energy interaction around the Fermi level, we only consider the coupling between the lowest phonon branch and the single electronic band which crosses the Fermi level [see the pink bands in Figs. S1(b) and S1(e)]. Due to the complexity in describing the EPC matrix elements, only static Lindhard susceptibility has been widely used to understand the CDW formation, while the q-EPC which however may play a more important role is ignored \cite{johannes2008fermi,zhu2015classification}. 

Here, to remedy this blemish, we applied density functional perturbation theory (DFPT) to obtain the accurate element $g$, which is given by \cite{giustino2017electron}: 

\begin{equation}
   g_{\mathbf{k},\mathbf{k}+\mathbf{q}}={(\frac{\hbar}{2M\omega_\mathbf{q}})}^{1/2}\bra{\varphi_{\mathbf{k}+\mathbf{q}}} \partial_\mathbf{q}V\ket{\varphi_\mathbf{k}}
\end{equation}

\noindent where $\partial_\mathbf{q}V$ is the derivative of the electron-ion potential with phonon frequency $\omega_\mathbf{q}$, $\varphi_\mathbf{k}$ is the electronic wavefunction with wavevector \textbf{k}. For a direct comparison, we also calculated the matrix element $g$ between electronic states \textbf{k} and \textbf{k+q} by the tight-binding (TB) model \cite{varma1979electron,flicker2015charge,flicker2016charge}, which is:

\begin{equation}
    g_{\mathbf{k},\mathbf{k}+\mathbf{q}}\propto\left(\mathbf{\nu}_\mathbf{k}-\mathbf{\nu}_{\mathbf{k}+\mathbf{q}}\right) \cdot \frac{\mathbf{q}}{|\mathbf{q}|}
\end{equation}

\noindent where $\mathbf{\nu}_\mathbf{k}$ is the electron velocity at \textbf{k} point in the coupled band, $\frac{\mathbf{q}}{|\mathbf{q}|}$ is the longitudinal projection as only LA phonons soften to zero. And this method succeeded in describing EPC properties in bulk 2H-NbSe$_2$ \cite{flicker2015charge,flicker2016charge} and monolayer 1T-VSe$_2$ \cite{chua2021coexisting}. To distinguish the two different sources of EPC matrix elements in the following discussions, the related quantities obtained by the TB method are denoted as ${\bar{g}}_\mathbf{q}^{TB}$ and $\chi_\mathbf{q}^{TB}$, while the quantities obtained by DFPT are used bare notation ${\bar{g}}_\mathbf{q}$ and $\chi_\mathbf{q}$.

\begin{figure*}  
\centering  
\includegraphics[width=12cm]{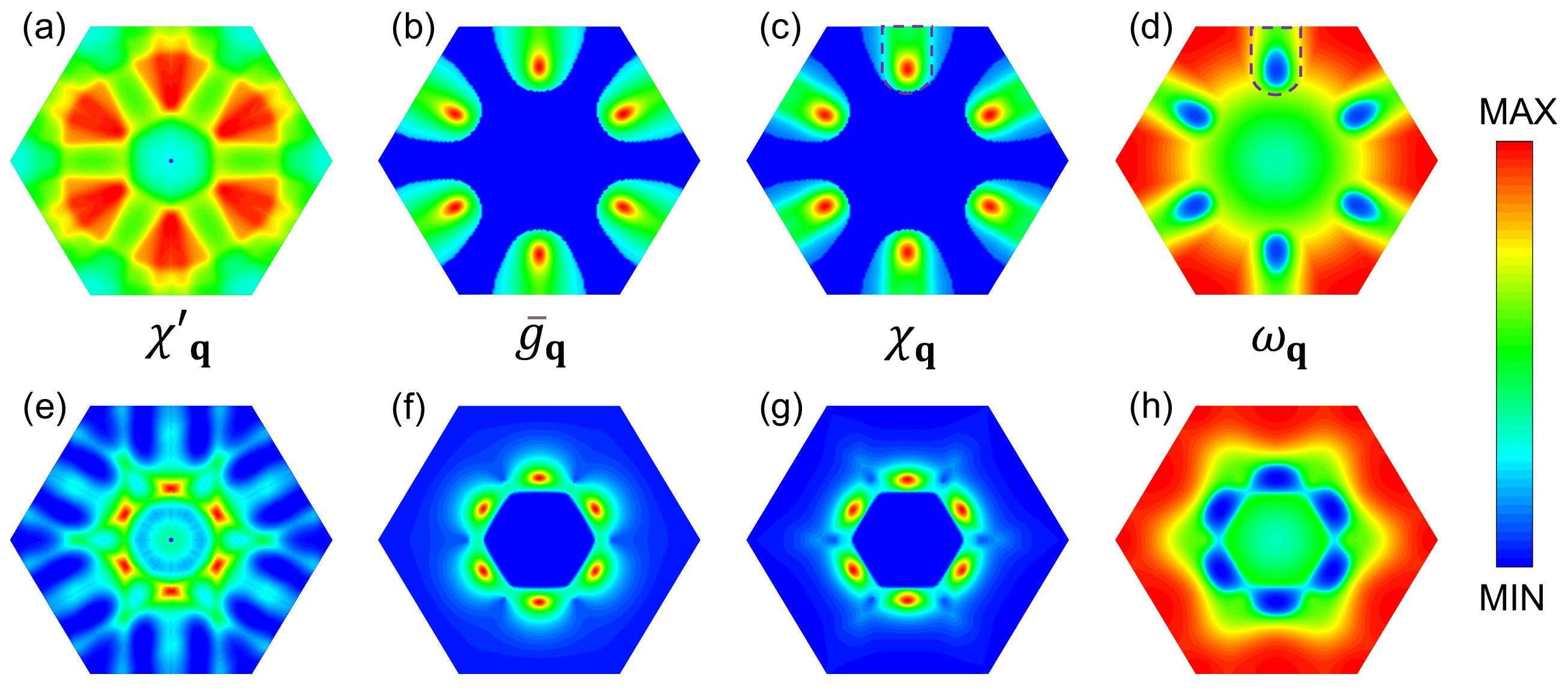}  
\caption{(a) Static Lindhard susceptibility ${\chi'}_\mathbf{q}$, (b) \textbf{q}-EPC ${\bar{g}}_\mathbf{q}$, (c) generalized static electronic susceptibility $\chi_\mathbf{q}$ and (d) the lowest phonon frequency $\omega_\mathbf{q}$ in the first BZ of NbSe$_2$. (e-h) Same as panel (a-d) but for VTe$_2$.
}  
\end{figure*}

The calculated ${\chi'}_\mathbf{q}$, ${\bar{g}}_\mathbf{q}$, $\chi_\mathbf{q}$, and $\omega_\mathbf{q}$ for NbSe$_2$ (upper panels) and VTe$_2$ (lower panels) are shown in Fig. 2. For NbSe$_2$, the static Lindhard susceptibility ${\chi'}_\mathbf{q}$ has a broad plateau from 2/5$\Gamma M$ to 4/5$\Gamma M$ [see Fig. 2(a) and the blue line in Fig. 3(a)], which indicates the weakness of FSN \cite{calandra2009effect}. Nonetheless, ${\bar{g}}_\mathbf{q}$ shows a strong electron-phonon interaction near 2/3$\Gamma M$ for NbSe$_2$ [see Fig. 2(b) and the red line in Fig. 3(a)]. In addition, the topology of $\chi_\mathbf{q}$ is very similar to ${\bar{g}}_\mathbf{q}$ [Figs. 2(b) and 2(c)] in the NbSe$_2$ first BZ, indicating the dominated role played by the \textbf{q}-EPC in $\chi_\mathbf{q}$. Fig. 2(d) displays the softened phonon modes of NbSe$_2$, which are concentrated in the arch-like area [see the purple dashed line in Fig. 2(d)], in agreement with the highland in $\chi_\mathbf{q}$ [see the purple dashed line in Fig. 2(c)]. More importantly, the peaks in ${\bar{g}}_\mathbf{q}$ and $\chi_\mathbf{q}$ [see the red areas in Figs. 2(b) and 2(c)] are at 2/3$\Gamma M$, which is consistent with the dip in $\omega_\mathbf{q}$ [see the blue areas in Fig.2(d)]. 

The EPC properties of bulk 2H-NbSe$_2$ have been well-described by using the TB method \cite{flicker2015charge,flicker2016charge}, however, we find this method fails in addressing monolayer 1H-NbSe$_2$. As for the generalized static electronic susceptibility $\chi_\mathbf{q}^{TB}$, the overall topology in the first BZ [Fig. S3(c)] cannot fit the phonon softening [Fig. 2(d)], which is in stark contrast to the good match achieved by DFPT [Fig. 2(c)]. Besides, the peaks of ${\bar{g}}_\mathbf{q}^{TB}$ and $\chi_\mathbf{q}^{TB}$ at the $\Gamma M$ path [see the green line in Fig. 3(a) and the orange line in Fig. 3(b)] are near 1/2$\Gamma M$, which predict the formation of $4\times4$ CDW instead of $3\times3$ CDW. By comparing to the prominent peaks of ${\bar{g}}_\mathbf{q}$ and $\chi_\mathbf{q}$ at 2/3$\Gamma M$ obtained by the DFPT [see the red line in Fig. 3(a) and the black line in Fig. 3(b)], we conclude that the \textit{ab-initio} based DFPT method is superior to the TB method in obtaining the \textbf{q}-EPC and predicting the CDW vector in monolayer NbSe$_2$.

\begin{figure} [!htb] 
\centering  
\includegraphics[width=8cm]{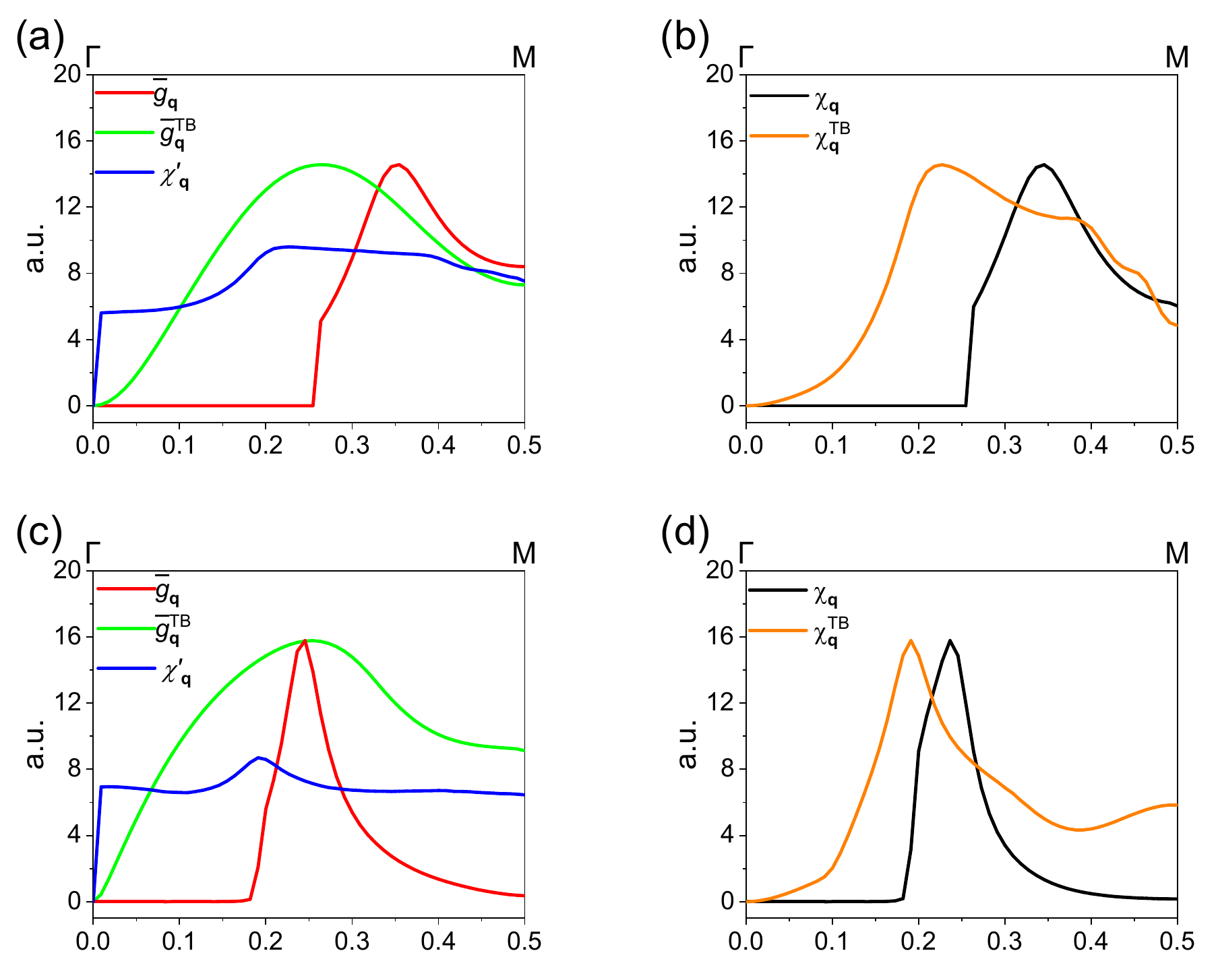}  
\caption{Direct comparison of the role of EPC and FSN along $\Gamma M$ path in NbSe$_2$ (up panel) and VTe$_2$ (down panel). (a) Comparing the difference among ${\bar{g}}_\mathbf{q}$ (red), ${\bar{g}}_\mathbf{q}^{TB}$ (green) and ${\chi'}_\mathbf{q}$ (blue) in NbSe$_2$, (b) Comparing the difference between $\chi_\mathbf{q}$ (black) and $\chi_\mathbf{q}^{TB}$ (orange) in NbSe$_2$. (c, d) Same as panels (a, b), but for VTe$_2$. Note that the unit of the y-coordinate is arbitrary units (a.u.) for straightforward comparison. 
}  
\end{figure}

\begin{figure*} [!htb]
    \centering
    \includegraphics[width=12cm]{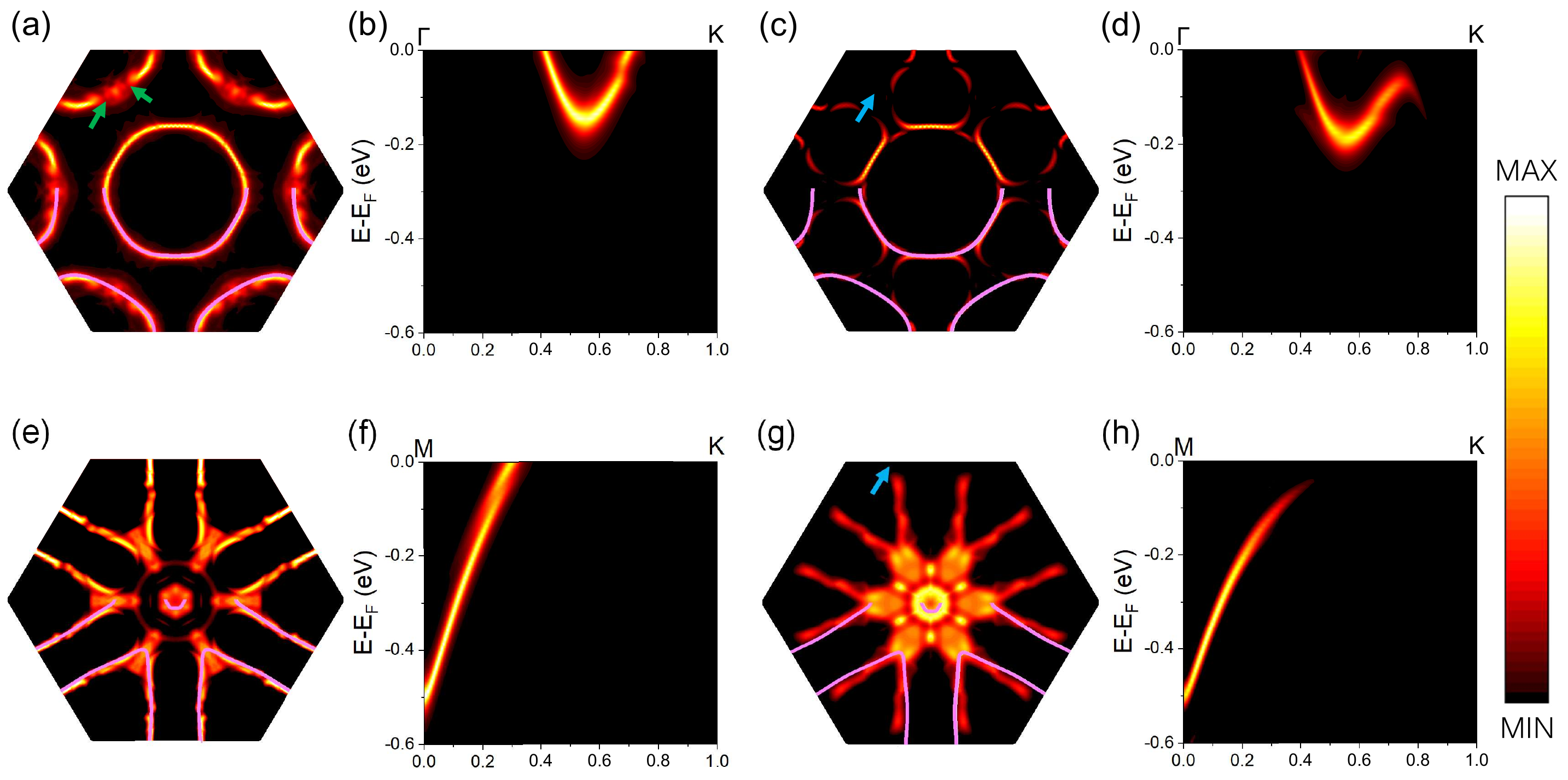}
    \caption{Simulation of the (a) Fermi surface and (b) the spectral function along $\Gamma K$ direction with constant $|g|$ of the NbSe$_2$ CDW system. (c, d) Same as (a, b), but with anisotropic $g$ in the simulation. Simulation of the (e) Fermi surface and (f) the spectral function along $MK$ direction with constant $|g|$ of the VTe$_2$ CDW system. (g, h) Same as (e, f), but with anisotropic $g$ in the simulation. The brightness of the dots denotes the spectral weights. The pink solid lines in (a, c, e, g) in the lower half of the BZ are the corresponding non-CDW Fermi surfaces for comparison. The green arrows in (a) indicate the partial CDW gaps opened by FSN, the blue arrows in (c, g) indicate the full CDW gaps opened by FSN + EPC. The paths of the spectral functions are chosen to better show the full CDW gaps.}
    \label{4}
\end{figure*}

For monolayer VTe$_2$, ${\chi'}_\mathbf{q}$ has a peak near 2/5$\Gamma M$ [see Fig. 2(e) and the blue line in Fig. 3(c)], in line with previous works \cite{wang2019evidence,sugawara2019monolayer}. However, this peak does not correspond to the $4\times4$ CDW structure observed from the experiments \cite{wang2019evidence,coelho2019monolayer}. Besides, the profile of $\chi_\mathbf{q}$ in the first BZ is again very close to ${\bar{g}}_\mathbf{q}$ [Figs. 2(f) and 2(g)], which both show maxima close to 1/2$\Gamma M$, providing a powerful clue of phonon softening at $\mathbf{Q^T}$. As shown in Fig. 2(h), the distribution of the softened phonon frequency $\omega_\mathbf{q}$ of VTe$_2$ shows hexapetalous flower-like pattern [see the blue area in Fig. 2(h)], which match well with the “hot” area in $\chi_\mathbf{q}$ [Fig. 2(g)]. One should note that the peak of ${\bar{g}}_\mathbf{q}$ is very sharp, which overwhelms the fluctuation of ${\chi'}_\mathbf{q}$ in VTe$_2$ [see the red line and the blue line in Fig. 3(c)], leading to the correct 1/2$\Gamma M$ peak position of $\chi_\mathbf{q}$ [see the black line in Fig. 3(d)]. However, the TB method still cannot explain the phonon softening in VTe$_2$. Although ${\bar{g}}_\mathbf{q}^{TB}$ shows a peak at 1/2$\Gamma M$, such peak is even broader than the nesting peak at 2/5$\Gamma M$, leading to the incorrect peak position of $\chi_\mathbf{q}^{TB}$ at 2/5$\Gamma M$ [see the green line in Fig. 3(c) and the orange line in Fig. 3(d)]. Compared the distribution of ${\bar{g}}_\mathbf{q}^{TB}$ [Fig. S3(e)], the “sharpness” of ${\chi'}_\mathbf{q}$ make the topology of $\chi_\mathbf{q}^{TB}$ closer to that of the ${\chi'}_\mathbf{q}$ in the first BZ of VTe$_2$ [Figs. 2(e) and S3(f)], which cannot explain the phonon softening shown in Fig. 2(h). Therefore, with the help of the accurate EPC matrix element $g$ obtained by DFPT, our study clearly demonstrates that it is \textbf{q}-EPC rather than FSN determines the phonon softening at correct positions and accounts for the CDW formation in both NbSe$_2$ and VTe$_2$.

\textit{Electron instabilities and CDW gaps.}--Having identified the domination of EPC in phonon softening, we now study the momentum-dependent CDW gap (electron instabilities) based on mean-field theory. The Hamiltonian of the CDW phase is minimally described by including one band crossing the Fermi level and electron-phonon interaction with phonon momentum \textbf{Q}. 

\begin{equation}
    H_{mf}=\sum_{\mathbf{k}}{\varepsilon_\mathbf{k}c_\mathbf{k}^\dag c_\mathbf{k}+\sum_{\mathbf{k},\mathbf{Q}}{2g_{\mathbf{k},\mathbf{k}+\mathbf{Q}}\Delta_\mathbf{Q}c_\mathbf{k}^\dag c_{\mathbf{k}+\mathbf{Q}}}}+h.c.
\end{equation}

\noindent Here, $c_\mathbf{k}^\dag$ ($c_\mathbf{k}$) and $\varepsilon_\mathbf{k}$ are creation (annihilation) operator and energy for an electron with momentum \textbf{k}. $\Delta_\mathbf{Q}$ is the order parameter which was approximated to a constant because of symmetry and small pocket size. This Hamiltonian can then be used to calculate the spectral function of the CDW phases [see SI-IV for more details about the theoretical background].

As shown in Fig. 4(a), the simulated Fermi surface of the NbSe$_2$ CDW structure with constant EPC matrix elements $|g|$ (i.e., $|g|=\sum_{\mathbf{k},\mathbf{k}+\mathbf{Q}} |g_{\mathbf{k},\mathbf{k}+\mathbf{Q}}|/{N_{\mathbf{k},\mathbf{k}+\mathbf{Q}}}$, where $N_{\mathbf{k},\mathbf{k}+\mathbf{Q}}$ is the number of $g$ in the calculation.) reflects the FSN effect under the mean-field picture. The norm of $g$ is used to avoid the arbitrary phase factor problem in the band basis of EPC matrix element. It is clearly shown that each K pocket has 3 couples of partial gaps [see the green arrows in Fig. 4(a)], where the spectral intensity becomes blurred as only partial electronic states are left at the Fermi surface. The partial gaps are at both sides of the $\Gamma K$ path, corresponding to the most heavily nested points of NbSe$_2$ [see the red points in white circle in Fig. S6(c)]. The incorporation of anisotropic matrix elements $g$ develop a more extensive gap opening on the Fermi surface [Fig. 4(c)], which considers the synergistic effects of FSN and EPC. Remarkably, the full band-gapped sectors can be found on the K pockets, where the electronic states on the Fermi surface are completely obliterated [see the blue arrow in Fig. 4(c)]. Furthermore, the spectral function was plotted along the $\Gamma K$ path, and we find the pure nesting effect cannot open a band gap along this path [Fig. 4(b)], in contrast with the spectral function derived with the anisotropic $g$, which obviously displays a full band gap close to the K point [Fig. 4(d)]. Considering there is no experimental report on the Fermi surface of the monolayer NbSe$_2$ CDW state, the predicted CDW gap distribution is compared with the unfolded Fermi surface of the simulated NbSe$_2$ CDW ground state, which displays a remarkable agreement with each other \cite{zheng2018first,zheng2019electron}. 

It is known that VTe$_2$ has triangular hole pockets, which has parallel sides to provide good nesting condition \cite{wang2019evidence,sugawara2019monolayer}. Such nesting will induce a peak in the static Lindhard susceptibility and possibly open a gap at the heavily nested point \cite{wang2019evidence}. In the phononic part discussion, FSN as the mechanism of the phonon softening in VTe$_2$ has been excluded, we now turn to discuss its relation to the CDW gap. Fig. 4(e) suggests that there is no obvious spectral weight depletion on the Fermi surface, and no CDW gap can be opened at the $MK$ path [Figs. 4(e) and 4(f)]. However, after considering the effect of the anisotropic EPC matrix elements $g$, there is a strong suppression of the spectral intensity near the M point [see the blue arrow in Fig. 4(g)]. No electronic state can be found on the Fermi surface at the $MK$ path, which indicates a full gap opening [Fig. 4(h)], consistent with the recent experimental results \cite{wang2019evidence}. Moving toward the $\Gamma$ point, the decreasing of the gap size is accompanied by the full to partial gap transition, and finally the gap is closed at the triangular K pocket apex [see Fig. 4(g)]. Such an anisotropic gap distribution on the Fermi surface agrees well with the angle-resolved photoemission spectroscopy (ARPES) measurements \cite{wang2019evidence}. 

\textit{Discussion.}--Although the CDW formation has been widely studied for decades, the underlying mechanism is still under debate. Using NbSe$_2$ and VTe$_2$ as examples, we explore the origin of their CDW orders by providing a comprehensive \textit{ab-initio} theoretical study. This study is vital because it is difficult to reconcile FSN and EPC so far. The main features of this work include an accurate description of the $\textbf{q}$-EPC in the whole BZ for the first time, correctly calculating the generalized static electronic susceptibility, the understanding of the CDW formation mechanism, and comparing the CDW gap distribution by the mean-field model with or without incorporating the EPC effects. All these results are self-consistent and emphasize the importance of EPC. 

We also address some puzzles surrounding the CDW properties of monolayer NbSe$_2$ and VTe$_2$. For monolayer NbSe$_2$, a small CDW gap of 4 meV is obtained by the molecular-beam epitaxy (MBE) grown samples with low $T_{CDW}\sim$ 25K \cite{ugeda2016characterization}. While the mechanically exfoliated samples show higher $T_{CDW}\sim$ 145K without a gap measurement \cite{lin2020patterns,xi2015strongly}, and it is believed that a large $T_{CDW}$ corresponds to a large gap size \cite{gruner1988dynamics,rossnagel2011origin,flicker2015charge,chen2018unique}. Previous experimental and theoretical studies indicate that the suppression of the $3\times3$ CDW order in the MBE-grown NbSe$_2$ samples is due to the charge transfer from the graphene substrate \cite{lin2020patterns,silva2016electronic,chen2020visualizing}. Therefore, the intrinsic NbSe$_2$ is expected to have a larger gap size than 4meV, as revealed by the previous unfolded band structure \cite{zheng2018first,lian2018unveiling} and our mean-field simulation. For monolayer VTe$_2$, previous work reported that there is no observable charge transfer between VTe$_2$ samples and the graphene substrate \cite{sugawara2019monolayer}, hence, the graphene based VTe$_2$ samples may be close to the freestanding one, which should have similar properties including the CDW gap opening. Our mean-field calculations successfully reproduce the experimental ARPES results as expected \cite{wang2019evidence}. Besides, the long parallel sides of the triangular hole pockets provide a good condition for FSN, which was thought to be the origin of the anisotropic CDW gap \cite{wang2019evidence}. However, our work excludes this hypothesis and emphasizes the significance of EPC in the CDW gap opening by providing a convincible description.

In conclusion, using first-principles calculations and the mean-field theory, we report a quantitative study of CDW properties in monolayer 1H-NbSe$_2$ and 1T-VTe$_2$. Our results confirm the validity of the EPC mechanism in a non-1D CDW system, which supplies previous experimental and theoretical works. The combined analysis of FSN and EPC in both phononic and electronic pictures constructs a profound understanding of the CDW formation mechanism. We argue that the same physics, in principle, should be applied to other higher-dimensional CDW systems. Besides, the proposed analytical method can be decorated by more accurate calculations (i.e., GGA+U, GW, hybrid function, etc.) \cite{giustino2017electron,zhou2021ab,li2019electron,yin2013correlation}, which allows further CDW studies in more complex systems. Our work paves a general way to unravel the physical insights of the CDW formation mechanism with phonon and electron instabilities, which can be extended for the understanding of charge ordering in other transition metal compounds \cite{manzeli20172d}, kagome metals \cite{tan2021charge} and high-temperature superconductors \cite{wang2020charge}.

~\\
\textit{Acknowledgement.}--
Z.W. thanks Dr. Felix Flicker for fruitful discussions. Z.W. also thanks Dr. Miguel Dias Costa and Yi-Ming Zhao for IT support. C.C. acknowledges the support from Shuimu Tsinghua Scholar Program. This research project is partially supported by the Ministry of Education, Singapore, under its MOE AcRF Tier 3 Award MOE2018-T3-1-002, and MOE AcRF Tier 2 Award MOE2019-T2-2-030. Numerical computations were supported by the National Supercomputing Center (NSCC) Singapore and Center of Advanced 2D Materials (CA2DM) HPC infrastructure.


\providecommand{\noopsort}[1]{}\providecommand{\singleletter}[1]{#1}%

\end{document}